\documentclass[aps,
prd,
english,
nofootinbib,
reprint]{revtex4-1}

\usepackage{amsfonts,amsmath,amssymb}
\usepackage{graphicx}
\usepackage[utf8]{inputenc}
\usepackage{hyperref}
\usepackage{babel}
\usepackage{enumitem}

\newcommand\e{{\rm e}}

\newcommand\be{\begin{equation}}
\newcommand\ee{\end{equation}}
\newcommand\bea{\begin{eqnarray}}
\newcommand\eea{\end{eqnarray}}

\begin{document}

\def\rhoo{\rho_{_0}\!} 
\def\rhooo{\rho_{_{0,0}}\!} 

\begin{flushright}
\phantom{
{\tt arXiv:2024.$\_\_\_\_$}
}
\end{flushright}

{\flushleft\vskip-1.4cm\vbox{\includegraphics[width=1.15in]{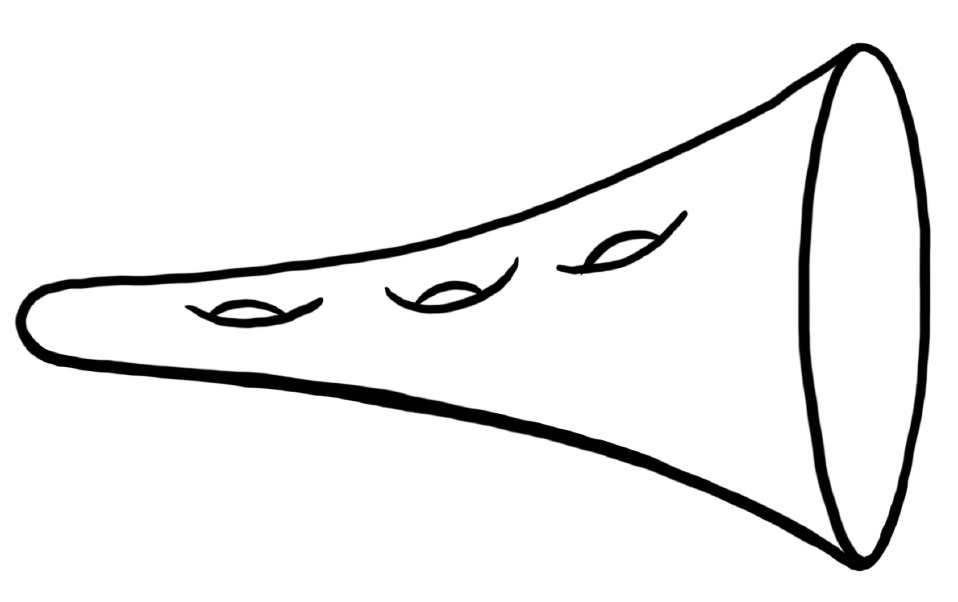}}}

\title{
Supersymmetric Virasoro Minimal Strings}
\author{Clifford V. Johnson}
\email{cliffordjohnson@ucsb.edu}

\affiliation{Department of Physics, Broida Hall,   University of California, 
Santa Barbara, CA 93106, U.S.A.}


\begin{abstract}
A  random matrix model definition of a family of ${\cal N}{=}1$ supersymmetric extensions of the Virasoro minimal string of Collier, Eberhardt, M\"{u}hlmann, and Rodriguez is presented. An analysis of the defining string equations shows that the models all naturally have unambiguous non-perturbative completions, which are explicitly supplied by the double-scaled  orthogonal polynomial techniques employed. Perturbatively, the multi-loop correlation functions of the model define a special supersymmetric class of ``quantum volumes'', generalizing the prototype case, some of which are computed. 
\end{abstract}


\maketitle



\section{Introduction}
\label{sec:introduction}
The Virasoro minimal string (VMS) recently presented in ref.~\cite{Collier:2023cyw} is an exciting new  kind of critical string theory.   It was shown to be perturbatively (in world-sheet topology) equivalent to a double-scaled~\footnote{The double-scaling limit\cite{Brezin:1990rb,*Douglas:1990ve,*Gross:1990vs,*Gross:1990aw} takes a large $N$ limit of an $N{\times}N$ matrix model while tuning parameters in its potential to a critical point, yielding a model of 2D gravity.} random matrix model whose leading density of states~$\rho^{(b)}_0(E)$ is   the universal Cardy density for states of energy $E$ in  a  2D conformal field theory (CFT). Here,~$b$ parameterizes the central charge of a spacelike Liouville component of the worldsheet theory: $c{=}1{+}6(b{+}b^{-1})^2$. What remains is a timelike Liouville component with ${\hat c}{=}1{-}6(b^{-1}{-}b)^2$ such that $c{+}{\hat c}{=}26$.  After  a field redefinition~\cite{Mertens:2020hbs,Fan:2021bwt}, this Liouville description becomes a model of 2D  gravity with a Jackiw-Teiteboim-like~\cite{Jackiw:1984je,*Teitelboim:1983ux} dilaton coupling (useful in the study of the near-horizon dynamics of black holes), but with a more general potential parametrized by $b$. Correlation functions of the VMS   were also shown~\cite{Collier:2023cyw} to be captured by computations in chiral 3D gravity. Furthermore, the presence of a timelike Liouville sector suggests that the target space of the string has a cosmological interpretation~\cite{Rodriguez:2023kkl,Rodriguez:2023wun}. 
These connections  unite several techniques and ideas from different areas of 
quantum gravity research, so it is likely that this type of construction will potentially lead to  new   illuminating results in  the program of quantum gravity.

A very natural next step (anticipated in ref.~\cite{Collier:2023cyw}, but not explored),  is to define  {\it supersymmetric} Virasoro minimal strings, starting with the ${\cal N}{=}1$ supersymmetric  density of states formula 
(Eq.~(\ref{eq:leading-spectral-density}) below). This paper will do just that, using a {\it fully non-perturbative} random matrix model approach,  capturing the  world--sheet topological expansion and much more besides. The output will be a large family of such string theories for each~$b$, depending upon the choice of the  matrix ensemble. (Note that there are other choices beyond those studied here~\footnote{\label{fn:class-B}Beyond the leading order equations to come, there will be a choice of matrix ensemble to be made. {\it E.g.}, the  models discussed here, of ``A-type'' built from single-cut $({\boldsymbol\alpha},2)$ AZ~\cite{Altland:1997zz} ensembles are distinct from a ``B-type'' that can be constructed as a Dyson class $\beta{=}2$ ensemble with the hard edge behaviour arising from two cuts merging. See refs.~\cite{Stanford:2019vob,Johnson:2021owr}, for the situation in  JT supergravity. These, and other choices  discussed in ref.~\cite{Stanford:2019vob} should be accessible as super-VMS models by varying~$b$ away from 1.}). All the models in this paper turn out to  be unambiguously non-perturbatively well-defined (this is only true for the $b{=}1$ case in the ordinary VMS). Moreover, various special features of the models  serve as predictions for the expected dual  3D gravity  and 2D dilaton gravity settings.

\section{The Approach}Following ref.~\cite{Collier:2023cyw}'s  presentation of the VMS,  ref.~\cite{Johnson:2024bue} formulated the random matrix description fully non-perturbatively by casting it into orthogonal polynomial language. There were two key steps:\\ {\bf (1)} Using input from the leading density of states to determine the  double-scaled potential of the matrix model by writing it as a sum of the appropriate basic multi-critical matrix models. This amounts to writing the leading ``string equation'': $\sum_{k=1}^\infty t_k u_0^k(x){+}x{=}0$
for  
specific numbers~$t_k$. Here $u_0(x)$ is the leading piece of $u(x)$,  the double-scaled orthogonal polynomial recursion coefficient, and~$x$ is the surviving part of what was (pre-double-scaling) the discrete  orthogonal polynomial index.~\footnote{Ref.~\cite{Castro:2024kpj} also independently completed  step~{\bf (1)} for  ordinary VMS.}\\  {\bf (2)} Determine and solve the relevant  non-linear ordinary differential equation (the complete ``string equation'') for~$u(x)$. The orthogonal polynomials themselves are then recovered as wavefunctions of an associated Schr\"odinger problem, with $u(x)$ as the potential.

This paper will show  how do the above steps to define an ${\cal N}{=}1$ class of supersymmetric VMS,  showing that the resulting equations defining the random matrix model give a compelling  definition string theories. Already at step~{\bf (1)} the leading order  string equation that results   turns out to have striking properties that protect it (for all $b$) from  pathologies that could have spoiled finding a non-perturbative completion. At step~({\bf 2}) the non-perturbative string equation describing the complete theory  then readily supplies perturbative topological  expansions for the physics, as well as the full non-perturbative physics of both orientable and non-orientable theories. 

\section{Recasting as Multicritical Models}
\label{sec:multicritical}
Beginning with the universal 
expression\footnote{See refs.~\cite{Kastor:1986ig,Matsuo:1986vc}, and {\it e.g.} refs.~\cite{Fukuda:2002bv,Ahn:2002ev,Mertens:2017mtv}. In a more recent context, Scott Collier and Henry Maxfield have an unpublished derivation of this formula, obtained from exploring crossing properties of ${\cal N}{=}1$ Virasoro characters, along the lines of ref.~\cite{Maxfield:2019hdt}.} for
the density of (NS-R) states with weight $h_p{=}\frac{(c{-}\frac32)}{24}{+}\frac{P^2}{2}{+}\frac{\delta}{16}$ 
in an ${\cal N}{=}1$ CFT and writing $E{=}P^2$ defines:
\be
\label{eq:leading-spectral-density}
\rho^{(b)}_{0}(E)  = \e^{S_0}{2\sqrt{2}}\frac{\cosh(2\pi b\sqrt{E})\cosh(2\pi b^{-1}\sqrt{E})}{\sqrt{E}}\ ,
\ee
with $0{\leq}b{\leq}1$ and  $c{=}\frac32{+}3Q^2$, $Q{=}b{+}b^{-1}$. The factor $\delta{=}1$ (or 0) in the Ramond (or Neveu-Schwarz) sector. Below, $\rho^{(b)}_{0}(E)$ will be engineered as the leading (disc) density of states of the world-sheet/2D-gravity model. Meanwhile, the ${\cal N}{=}1$ critical string construction analogous to that of the ordinary Virasoro minimal string has~\cite{Collier:2023cyw} a spacelike (super)-Liouville with $c$, $h_P$,  and $Q$ as defined above, and also a (timelike) (super)-Liouville with  ${\hat c}{=}\frac32{-}3{\widehat Q}^2$, ${\widehat Q}{=} b^{-1}\!-\!b$ and ${\hat h}_{\widehat P}$, such that $h_P{+}{\hat h}_{\widehat P}{=}1$ and $c{+}{\hat c}{=}15$.

In Eq.~(\ref{eq:leading-spectral-density}), $\hbar{=}{\rm e}^{-S_0}$ is the topological expansion parameter, and  $S_0$ is an extremal entropy in the dilaton  gravity picture. 2D Euclidean world-sheets with Euler characteristic $\chi{=}2{-}2g{-}n$ (where $g$ counts handles and~$n$ counts boundaries) come with a factor $\hbar^{-\chi}$. Below, $\hbar$ is identified with the (renormalized) $1/N$ expansion parameter of the double-scaled matrix model. 

Notice that the low energy tail of $\rho_0^{(b)}(E)$ is~${\sim}1/\sqrt{E}$, characteristic of a ``hard edge" matrix model. This is consistent with the fact that the lowest energy of a supersymmetric system is $E{=}0$, and so in the random matrix description there should be a hard wall there that stops energy eigenvalues from flowing to  $E{<}0$. Furthermore,  when $b{\to}0$, $\rho_0^{(b)}(E)$  becomes (after a rescaling of~$E$) the leading spectral density of ${\cal N}{=}1$  JT supergravity, for which this property has already been incorporated into  a random matrix model description, perturbatively~\cite{Stanford:2019vob} and non-perturbatively~\cite{Johnson:2020heh,Johnson:2020exp,Johnson:2021owr}. Among  the hard edge models used were   random matrix models of type $(\boldsymbol{\alpha},\boldsymbol{\beta})$ in the Altland-Zirnbauer (AZ)~\cite{Altland:1997zz} classification. The ensemble type  should not change as~$b$ varies, and so these will the kinds of model sought here, with cases  $\boldsymbol{\beta}{=}2$ and $\boldsymbol{\alpha}{=}0,1$ or~2  as the focus. (Footnote~2 has other options.)

If $\rho_b^{(b)}(E)$ is indeed  a  double-scaled random matrix model's leading spectral density,  it is to be expected that: 
\be
\label{eq:spectral-density-leading}
\rho^{(b)}_{0}(E) \!= \frac{1}{2\pi\hbar}\int_{-\infty}^\mu\!\frac{\Theta(E{-}u_0(x)) dx}{\sqrt{E-u_0(x)}}=\frac{1}{2\pi\hbar}\int_{0}^E\!\frac{f(u_0)du_0}{\sqrt{E-u_0}} \ ,
\ee where the previously mentioned function $u_0(x)$ satisfies $\sum_{k=1}^\infty t_k u_0^k(x){+}x{=}0$,
$f(u_0){=}{-}\partial x/\partial u_0$  and the $t_k$ and~$\mu$ are  determined by this matching. By expanding both sides in $E$, or by directly inverting the integral transform~\cite{Johnson:2020lns}, some algebra yields the pleasing result:
\be
\label{eq:new-teekay}
t_k=2\sqrt{2}\pi\frac{\pi^{2k}}{(k!)^2}
\left(Q^{2k}+{\widehat Q}^{2k}\right)\ ,\quad \mu=t_0=4\sqrt{2}\pi\ ,
\ee
and so the leading string equation can be written as:
\bea
2\sqrt{2}\pi\left[I_0(2\pi Q\sqrt{u_0}\,)+I_0(2\pi{\widehat Q}\sqrt{u_0}\,)\right]+x=0\ .
\label{eq:leading_string_equation}
\eea
Here $I_0(y)$ is  the modified Bessel  function in $y$ of order~$0$. 
Comparison with the results derived in ref.~\cite{Johnson:2024bue} (and independently in ref.~\cite{Castro:2024kpj}) for the ordinary VMS will reveal a close similarly. In fact the {\it only} adjustment is that while there was a {\it difference} between pieces involving $Q$ and those involving~$\widehat{Q}$, now there is a {\it sum}! 

\begin{figure}[t]
\centering
\includegraphics[width=0.49\textwidth]{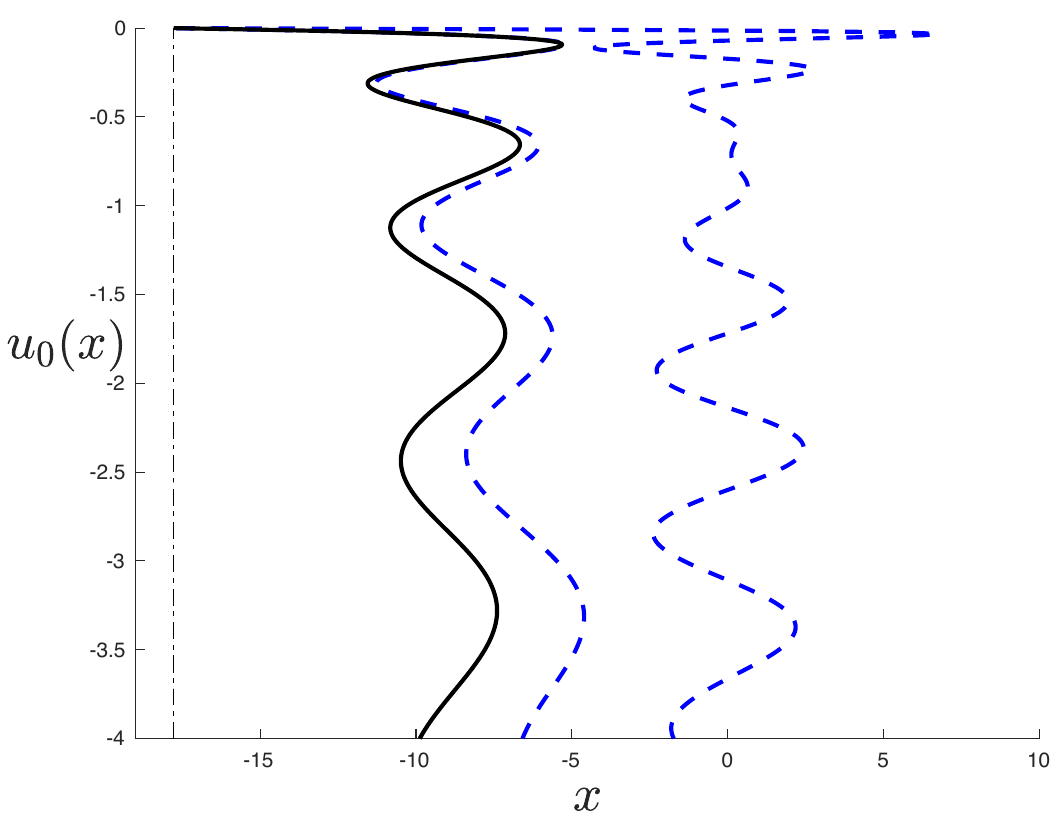}
\caption{\label{fig:u0-examples} The potential  $u_0(x)$ showing the (harmless) multi-valuedness for the $b{=}1, 0.95$, and $0.3$ cases. The dot-dash vertical line to the left is $x{=}{-}4\sqrt{2}\pi$, past where where $u_0(x)$ rises to be positive and single-valued. 
}
\end{figure}

The next step is to check whether there is any problematic multi-valuedness of $u_0(x)$ that would imply~\cite{Johnson:2020lns,Johnson:2021tnl} semi-classical obstructions to finding an unambiguous non-perturbative definition,  as was done in the ordinary VMS case~\cite{Johnson:2024bue,Castro:2024kpj}. There, the only case that survived was $b{=}1$, where the multi-valuedness present was all at $x{>}0$, outside the integration region used to define the density (and other matrix model quantities). Undulations from both $Q$ and~$\widehat{Q}$ sectors (arising when $b{<}1$) inevitably spilled into~$x{<}0$. Here, it all looks generically to be problematic again, since those sectors both contribute again for general~$b$. However something rather excellent happens: $u_0(x)$ vanishes at $x{=}{-}4\sqrt{2}\pi$ instead of zero, and for all $b$ the undulations do not stretch to smaller $x$ than this. To establish this, note that in this $u_0{<}0$ region, the case $b{=}1$ is the curve $2\sqrt{2}\pi\left[J_0(2\pi\sqrt{-u_0})+1\right]{+}x{=}0$, which vanishes at $x{=}{-}4\sqrt{2}\pi$, and never returns to this value of $x$ because of the decaying nature of Bessel undulations. For other~$b$, the undulations are pushed further to the right by a purely additive contribution. 
Since the~$x$ integral in the $x{<}0$ region should stop where $u_0(x){\to}0$, which is $x{=}{-}4\sqrt{2}\pi$, all cases have a harmless multi-valuedness. The  relevant parts of $u_0(x)$ with three cases of $b$ is shown in Fig.~\ref{fig:u0-examples}. 

Meanwhile, for the $x{>}0$ regime, a different solution is used, as is familiar from analogous cases studied in the JT supergravity context starting with refs.~\cite{Johnson:2020heh,Johnson:2020exp}. It is simply $u_0(x){=}0$ from $x{=}0$ all the way to $\mu{=}{+}4\sqrt{2}\pi$. That this is correct is consistent with the fact that the key leading term $2\sqrt{2}/(\hbar\sqrt{E})$  in the spectral density  is generated by the right hand side of relation~(\ref{eq:spectral-density-leading}) as $\mu/(2\pi\hbar\sqrt{E})$. It is surprising that there is a jump in the $x$-integration (from $-4\sqrt{2}\pi$ to 0) in order for everything to work so nicely. This mirrors a surprising such jump discovered recently~\cite{Johnson:2023ofr} in the context of ${\cal N}{=}2$ JT supergravity.\footnote{The lesson seems to be that for supersymmetric cases (where part of the defining $x$-integral always runs to $x{>}0$), previous examples where the integral in the $x{<}0$ regime ends at $x{=}0$ were simply special cases. Moreover the ${\cal N}{=}2$ case with non-zero threshold energy~\cite{Johnson:2023ofr} for the non-BPS sector showed that the integral in the $x{>}0$ sector needn't even {\it start} at $x{=}0$.} 

Step {\bf (1)} has now been completed, and it has been shown that the leading string equation that results for $u_0(x)$ for all $b$  nimbly avoids showing any semi-classical avatars of non-perturbative problems. It is time to see how this all fits into a complete 
matrix model definition.

\section{The full non-perturbative definition}
\label{sec:fully-nonp}
Recall that the topological expansion parameter is~$\hbar{=}\e^{-S_0}$, and for the closed string sector $u(x){=}u_0(x){+}\sum_{g=1}^\infty u_{2g}(x)\hbar^{2g}{+}\ldots$ where the ellipses denote non-perturbative parts. Actually, $u(x)$ is  the second derivative of the  partition function~${\widetilde Z}(\mu)$ of the string theory, nicely expressed through:
\be
{\widetilde Z} = \hbar^{-2}\int_{-\infty}^\mu (x-\mu)u(x) dx=\sum_{g}{\widetilde Z}_{2g}\hbar^{2g-2}+\cdots\ ,
\ee
defining the topological expansion for the world-sheets.

For the ensembles that are the focus here, the appropriate non-linear ordinary differential equation for defining the full function $u(x)$ is:
\be
\label{eq:big-string-equation}
u{\cal R}^2-\frac{\hbar^2}2{\cal R}{\cal R}^{\prime\prime}+\frac{\hbar^2}4({\cal R}^\prime)^2=\hbar^2\Gamma^2\ ,
\ee
with ${\cal R}{\equiv}\sum_{k=1}^\infty t_k R_k[u]{+}x$ where the $R_k[u]$ are polynomials (see below) in $u(x)$ and its $x$-derivatives. 
This ODE 
arose in early studies~\cite{Dalley:1991qg,*Dalley:1992br,*Dalley:1991vr,*Morris:1991cq,*Dalley:1991xx,*Anderson:1991ku} of ensembles of {\it positive} matrices $M$  (later identified to be of type  $(2\Gamma{+}1,2)$ in the AZ classification~\cite{Altland:1997zz}).   The $ R_k[u]$ are the ``Gel'fand-Dikii''~\cite{Gelfand:1975rn} differential polynomials  in $u(x)$ and its derivatives, normalized here so that the non-derivative part has unit coefficient: $R_k{=}u^k+\cdots+\#u^{(2k-2)}$ where $u^{(m)}$ means the $m$th $x$-derivative. {\it e.g.}, $R_1{=}u$, $R_2{=}u^2{-}\frac{\hbar^2}{3}u^{\prime\prime}$, $R_3 {=} u^3+{\hbar^2}(u^\prime)^2/2+{\hbar^2}uu^{\prime\prime}+{\hbar^4}u^{(4)}/{10}$, {\it etc.}

The solutions needed here are specified by boundary conditions that can be written for $\hbar{\to}0$, where the string equation becomes $u_0{\cal R}_0^2=0$, with ${\cal R}_0{\equiv}\sum_k t_k u_0^k+x$: 
\bea 
u_0=0 \,\,\, \mbox{as}\,\,\, x\to+\infty\ , \,\,\mbox{and} \,\,
 {\cal R}_0=0 \,\,\, \mbox{as}\,\,\, x\to-\infty\ .\label{eq:boundary-conditions}
 \eea
 The $t_k$   have been  determined  in Eq.~(\ref{eq:new-teekay}), and so the definition (and hence step {\bf (2)}) is complete. Different values of  $\Gamma$ correspond to distinct random matrix ensembles. Both integer and half-integer~$\Gamma$ are allowed, and $\Gamma{=}{\pm}\frac12$ and $0$ will be the focus. It is  natural to declare that they each define a {\it distinct} kind of ${\cal N}{=}1$ supersymmetric Virasoro minimal string. This is consistent with the $b{\to}0$ limit which yields a known~\cite{Stanford:2019vob} family of  JT supergravity models. Just as in that limit, the  cases $\Gamma{=}{\pm}\frac12$ are very special, as will be discussed. Note that they are unorientable theories, while $\Gamma{=}0$ is orientable.


 

\section{Non-Perturbative Results}
\label{sec:non-pert}
The next step is to solve  Eq.~(\ref{eq:big-string-equation}) for $u(x)$ with the boundary conditions~(\ref{eq:boundary-conditions}). 
The string equation is formally of infinite order, since each $t_k$ controls a term with $2k{-}2$ derivatives and all the~$t_k$ are turned on ($k{=}1,\ldots,\infty$). However ref.~\cite{Johnson:2020exp} noted that since the $t_k$ decrease swiftly enough in size as~$k$ increases, a sensible truncation of the equation can be done that can capture the physics up to any desired accuracy.
\begin{figure}[t]
\centering
\includegraphics[width=0.49\textwidth]{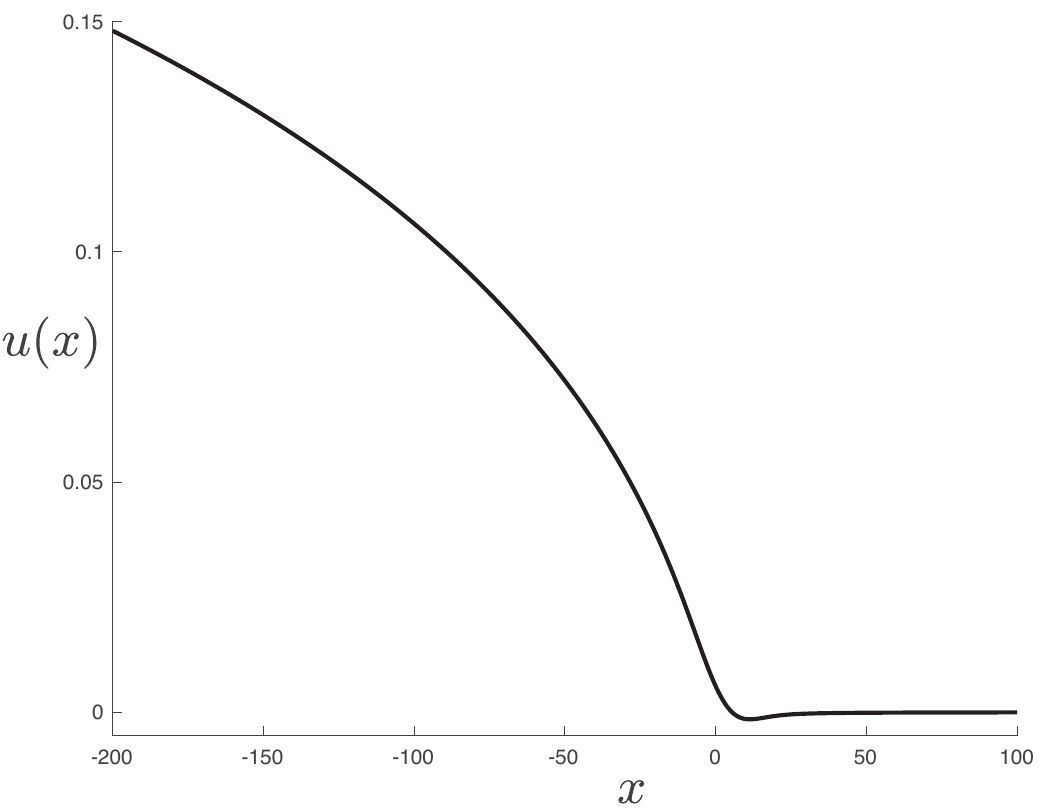}
\caption{\label{fig:full-potential} The potential  $u(x)$ for the $\Gamma{=}1$, $b{=}1$ case.   Here $\hbar{=}{\rm e}^{-S_0}{=}1$. }
\end{figure}
Those methods were used here to generate example solutions. Fig.~\ref{fig:full-potential} shows the case for  $b{=}1$, $\Gamma{=}0$.

\begin{figure}[t]
\centering
\includegraphics[width=0.48\textwidth]{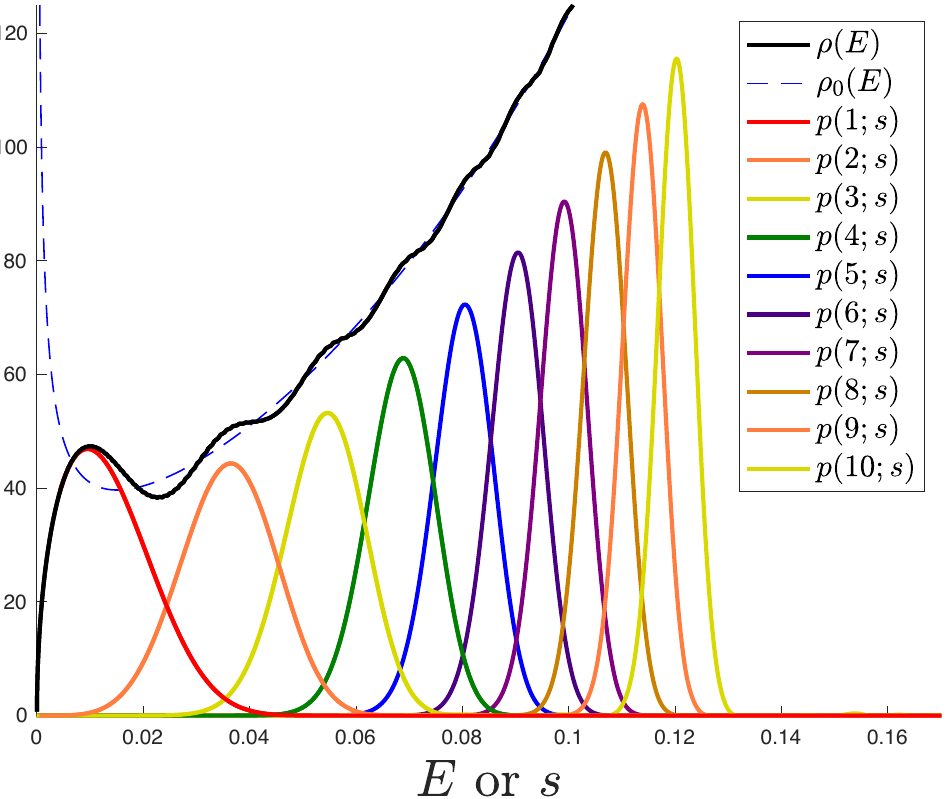}
\caption{\label{fig:full-spectral-density-B} The full spectral density $\rho(E)$ for the $\Gamma{=}\frac12$, $b{=}1$ Supersymmetric Virasoro minimal string. The dashed line is the disc level result of Eq.~(\ref{eq:leading-spectral-density}).  Also shown are ten of the microstate probability distributions $p(m,s)$. Here $\hbar{=}{\rm e}^{-S_0}{=}1$. }
\end{figure}

As mentioned above, $u(x)$ is the scaling limit of the recursion coefficient from which all the orthogonal polynomials can be determined, and $x$ labels the index on the orthogonal polynomials, now itself a continuous coordinate in the large $N$ scaling limit. The orthogonal polynomials themselves   become functions $\psi(E,x)$ in the limit, where $E$ is the scaling piece of the (continuous) eigenvalue coordinate $\lambda$ near the end of the leading distribution.   Once $u(x)$ is known, it turns out that the 
 $\psi(E,x)$ are determined (up to  normalization) as wavefunctions of a Schr\"odinger problem with $u(x)$ as the potential: 
\be 
\label{eq:schrodinger-problem}
\left[-\hbar^2\frac{\partial^2}{\partial x^2}+u(x)\right]\psi(x,E)=E\psi(x,E)\ ,
\ee
 Many things can be computed in  the random matrix model  with the $\psi(E,x)$. A fundamental object  is:
\be
\label{eq:kernel}
K(E,E^\prime) = \int_{-\infty}^\mu\!\!  dx \, \psi(E,x)\psi(E^\prime,x) \ ,
\ee
a kernel whose  derivation and uses are reviewed in this context in ref.~\cite{Johnson:2022wsr}.
Knowing it fully non-perturbatively ({\it via}  the $\psi(E,x)$) allows for physics that is entirely   inaccessible in topological perturbation theory to be probed. 
%
%
The diagonal of $K(E,E^\prime)$ is the spectral density:
\be
\label{eq:spectral-density-exact}
\rho(E) =\int_{-\infty}^\mu \left|\psi(x,E)\right|^2 dx\ . 
\ee
Its leading piece, met earlier in Eq.~(\ref{eq:spectral-density-leading}) comes from using the leading WKB form of  $\psi(E,x)$ in the $\hbar{\to}0$ limit.
\begin{figure}[t]
\centering
\includegraphics[width=0.48\textwidth]{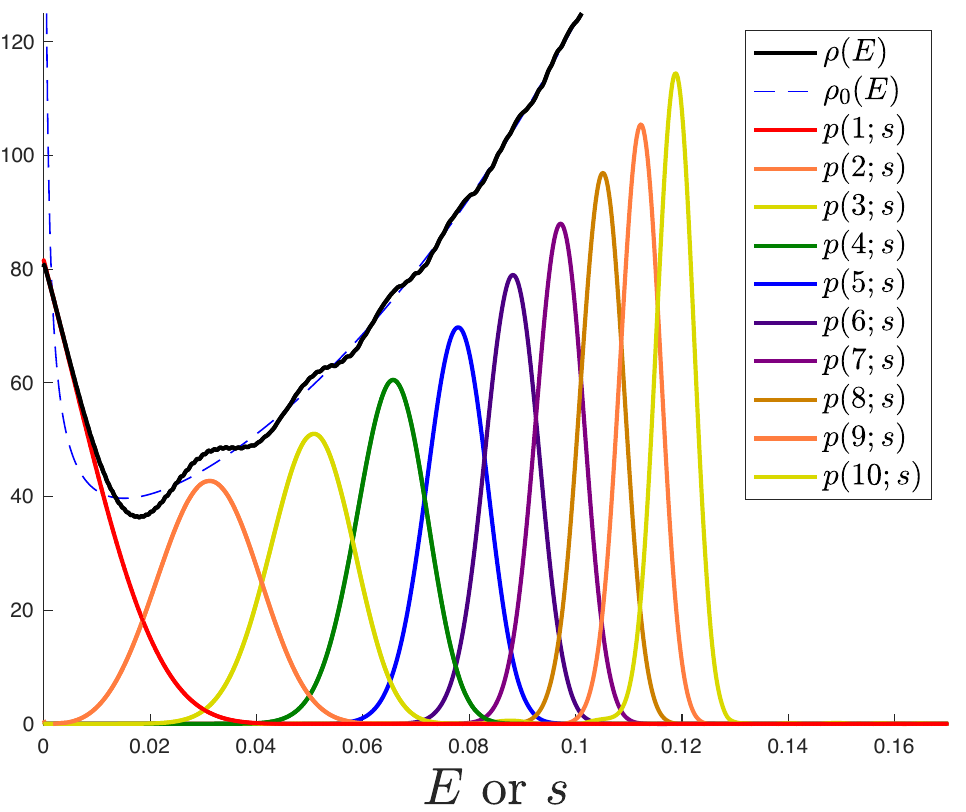}
\caption{\label{fig:full-spectral-density-A} The full spectral density $\rho(E)$ for the $\Gamma{=}0$, $b{=}1$ Supersymmetric Virasoro minimal string. The dashed line is the disc level result of Eq.~(\ref{eq:leading-spectral-density}).  Also shown are ten of the microstate probability distributions $p(m,s)$. Here $\hbar{=}{\rm e}^{-S_0}{=}1$. }
\end{figure}
As an example of these computations,  $\rho(E)$ is shown as top the dark solid  lines in Figs.~\ref{fig:full-spectral-density-A} and~\ref{fig:full-spectral-density-B} (for the cases of $\Gamma{=}0$, and $\Gamma{=}\frac12$, with  $b{=}1$). The dashed line is the leading  spectral density~(\ref{eq:leading-spectral-density}) for $b{=}1$. Also shown is an especially  intrinsically non-perturbative structure computable with $K$ (as kernel in a Fredholm problem): The  set of curves giving the probability   $p(m,s)$ that  the  $m$th eigenvalue of a matrix in  the ensemble takes the value~$s$. This is  information about the underlying discrete microscopic physics of the 2D quantum gravity~\cite{Johnson:2021zuo}, recasting the continuous spectral density as a {\it discrete} sum of peaks:
$\rho(E) {=} \sum_{m=1}^\infty p(m,E).$

\section{Remarks on Geometry and Perturbation Theory}
\label{sec:geometry}
An important observable is the Euclidean 2D gravity 
 partition function:
$Z(\beta){\equiv}\langle{\rm Tr}({\rm e}^{-\beta M})\rangle{=}\int \rho(E){\rm e}^{-\beta E} dE$,
where~$M$ is the matrix, thought of in the string context as a particular kind of loop operator with   length $\beta$. In fact, for the Virasoro minimal string, on a surface of genus $g$ the $n$--point correlation function $\langle Z(\beta_1),\ldots,Z(\beta_n)\rangle_{g,n}$ is built (generalizing the construction in ref.~\cite{Saad:2019lba} involving trumpet factors) from quantities ${\widehat V}^{(b)}_{g,n}(P_1,\ldots, P_n)$, (where~$P_i$ are Liouville momenta setting the length of geodesic boundaries). 
These are referred to in ref.~\cite{Collier:2023cyw} as  ``quantum volumes'', generalizing the Weil-Petersson volumes on the moduli space of bordered hyperbolic Riemann surfaces that are recovered in the classical limit $b{\to}0$ that yields JT gravity. Analogous correlation functions of the ${\cal N}{=}1$ VMS constructed here naturally define  supersymmetric variants of the quantum volumes, and the $b{\to}0$ limit should connect them to the  Weil-Petersson volumes of the super-Riemann surfaces of ${\cal N}{=}1$  JT supergravity~\cite{Stanford:2019vob,Norbury:2017eih}. The underlying recursive structures arising in the matrix model  yield recursion relations among these new volumes. How this works for the case of (super) ${\widehat V}^{(b)}_{g,1}(P_1)$ is explored  in Appendix~\ref{sec:one-boundary-multiple-handles}, with examples.  Here, the special case of genus zero is examined.

For $g{=}0$ the matrix model computation of the correlators for $n{\geq}2$ can be written 
as~\cite{Ginsparg:1993is,Ambjorn:1990ji,Moore:1991ir} (here $\beta_T{=}\sum_i\beta_i$):
\be
\label{eq:n-point-correlator}
\hskip-0.3cm\langle \prod_i Z(\beta_i)\rangle_{0,n}\simeq \frac{\sqrt{\beta_1\cdots\beta_n}}{2\pi^{n/2}\beta_T}\left[ (\hbar\partial_x)^{n-2}{\rm e}^{-\beta_T u_0(x)}
\right]_{x=\mu}\ ,
\ee
where everything is evaluated at the upper $x$-boundary, $x{=}\mu$.
It is worth making a qualifying remark here, that will also apply to the results of Appendix~\ref{sec:one-boundary-multiple-handles}. Given that in the leading perturbation theory it seems that a gap opens up in the integration region from $x{=}{-}4\sqrt{2}\pi$ to $x{=}0$ (see the discussion below Eq.~(\ref{eq:leading_string_equation})), one might worry that there could be additional contributions to correlators from these boundaries too. The general treatment of such regions is unclear at present, so this could be a possibility. However, the results should connect, in the $b{\to}0$ limit, to known results for ${\cal N}{=}1$ JT supergravity, and  it is unclear how such additional contributions would disappear in that limit (where no such contributions are present). Therefore the simple assumption that only the $x{=}\mu$ boundary contributes will be tentatively adopted here.

 Since $u_0(x)$ and its derivatives vanish at $x{=}\mu$ (see Eq.~(\ref{eq:boundary-conditions})), the correlators~(\ref{eq:n-point-correlator}) all vanish at leading order for these super VMS models.\footnote{In contrast, while $u_0$ vanishes at $x{=}{-}4\sqrt{2}\pi$, its derivatives do not,  which is why that edge of the interior integration region is of potential concern, given the remarks of the previous paragraph.}  Higher $g$  corrections are generated by  $u(x)$'s perturbative corrections, $u_{2g}(x)$, which are all seeded by this leading form of the solution of Eq.~(\ref{eq:big-string-equation}) in the $x{>}0$ region: $u(x){=}0{+}\hbar^2(\Gamma^2{-}\frac14)/x^2+\cdots$. 
For the cases $\Gamma{=}{\pm}\frac12$, this nicely suggests a key result, generalizing a similar JT supergravity story\cite{Stanford:2019vob,Johnson:2020heh,Johnson:2021tnl}: The $n{>}2$ correlators would seem to all vanish at every order in perturbation theory  because {\it all} the   $u_{2g}(x)$ recursively inherit the $(\Gamma^2{-}\frac14)$ factor and hence vanish  for $x{>}0$, where the derivatives are evaluated in~Eq.~(\ref{eq:n-point-correlator}). The same vanishing must be  true for the ${\cal N}{=}1$ quantum volumes for $n{>}2$. This is   a clear prediction for any  chiral 3D gravity and intersection theory setting where these correlators are computed, or indeed in the related ${\cal N}{=}1$ dilaton gravity scenario.

\section{Closing Remarks}
\label{sec:closing}
This paper presents a non-perturbative definition of  a family of  supersymmetric generalizations of ref.~\cite{Collier:2023cyw}'s  Virasoro minimal string. It seems likely that there is a (super) Liouville presentation of these models,  as well as a map to a  dilaton  supergravity on the one hand, and to 3D gravity and intersection theory on the other, at least perturbatively. These would be interesting to make explicit.  The models' excellent non-perturbative behaviour for all~$b$ should make them  especially sharp tools for quantum gravity and string theory in all these settings, and perhaps beyond.

 \section*{acknowledgments}
 
CVJ  thanks  the  US Department of Energy for support (under award \#\protect{DE-SC} 0011687), Scott Collier and Henry Maxfield for conversations, the KITP ``What is string theory?'' program (with partial support provided by The National Science Foundation Grant No. NSF PHY-1748958), and  Amelia for her support and patience.    


\onecolumngrid

\appendix
\section*{Appendices}
\section{One boundary, multiple handles}





\label{sec:one-boundary-multiple-handles}
Building on observations in ref.~\cite{Johnson:2020heh}, ref.~\cite{Johnson:2024bue} explored how to recursively compute the ``quantum volumes'' (in the language of ref.~\cite{Collier:2023cyw}) ${\widehat V}^{(b)}_{g,1}(P_1)$ for the ordinary Virasoro minimal string, using the fact that they (or rather their Laplace transform $\omega^{(b)}_{g,1}(z_1)$) are essentially contained in an ordinary differential equation~\cite{Gelfand:1975rn}:
\be
4(u(x) - E){\widehat R}^2-2\hbar^2{\widehat R}{\widehat R}^{\prime\prime}+\hbar^2({\widehat R}^\prime)^2 = 1\ ,
\ee
where ${\widehat R}(x,E){\equiv}<\!\!x| ({\cal H}{-}E)^{-1}|x\!\!>$ is the Gel'fand-Dikii resolvent of the Schr\"odinger operator~(\ref{eq:schrodinger-problem}). The equation was solved recursively order by order in $\hbar^2$, using as input the asymptotic  expansion  $u(x){=}u_0(x)+\sum_{g=0}^\infty u_{2g}(x)\hbar^{2g}{=}\cdots$, coming from the defining string equation of the model. This gives an expansion:
\bea 
\label{eq:gelfand-dikii-A}
{\widehat R}(x,E) &=& -\frac{1}{2}\frac{1}{[u_0(x)-E]^{1/2}}+\sum_{g=1}^\infty {\widehat R}_g(x,E)h^{2g}+\cdots\ \nonumber\\
&&\hskip-1.0cm= -\frac12\frac{1}{[u_0(x)-E]^{1/2}}
+
\frac{\hbar^2}{64}\left\{
\frac{16u_2(x)}{[u_0(x)-E]^{3/2}}
+\frac{4u^{\prime\prime}_0(x)}{[u_0(x)-E]^{5/2}}-\frac{5(u^{\prime}_0(x))^2}{[u_0(x)-E]^{7/2}}\right\}
\nonumber\\
&&\hskip-0.3cm+\frac{\hbar^4}{4096}\left\{
\frac{1024u_4(x)}{[u_0(x)-E]^{3/2}}
-\frac{256[3u_2(x)^2-u^{\prime\prime}_2(x)]}{[u_0(x)-E]^{5/2}}
+\frac{64[u^{(4)}_0(x) -10u_2(x) u^{''}_0(x)-10u^\prime_0(x)u^\prime_2(x)]}{[u_0(x)-E]^{7/2}}\right. \nonumber \\
&&\hskip-0.2cm\left.
-\frac{16[28u^{(3)}_0(x)u^\prime_0(x)+21u^{\prime\prime}_0(x)^2-70u_2(x)u_0(x)^2]}{[u_0(x)-E]^{9/2}}
+\frac{1704u^{\prime}_0(x)^2u^{\prime\prime}_0(x)}{[u_0(x)-E]^{11/2}}-\frac{1155u^{\prime}_0(x)^4}{[u_0(x)-E]^{13/2}}\right\} +\cdots\ ,
\eea
where one order higher has been included here as compared to ref.~\cite{Johnson:2024bue} for reasons to become clear below.
As a last step,  writing $z_1^2{=}{-}E$, the $\omega_{g,1}(z_1)$ were identified with (after multiplying by $(-2z_1)$) the quantities $\int^\mu {\widehat R}_g(x,E) dx$.

In that case of the  ordinary VMS, perturbation theory was developed entirely in the $x{<}0$ regime, and there $\mu{=}0$. In the  supersymmetric VMS case defined in the current work, perturbation theory beyond the disc comes entirely from  the $x{>}0$ region, with $\mu{=}4\sqrt{2}\pi$. This is where attention must be focused (but keeping in mind the caveat about interior boundaries  of the $x$ integration mentioned below Eq.~(\ref{eq:n-point-correlator}) and in footnote 6). 

Several simplifications occur, following from the fact that after a little algebra, the string equation~(\ref{eq:big-string-equation}) yields a very simple structure for the leading few terms of its $x{>}0$ expansion for $u(x)$, starting with the vanishing of $u_0(x)$:
\be
u(x)=0+\left(\Gamma^2-\frac14\right)\frac{\hbar^2}{x^2}-2t_1\left(\Gamma^2-\frac14\right)\left(\Gamma^2-\frac94\right)\frac{\hbar^4}{x^5}+\cdots
\ee
Notice the appearance of $t_1$ in the $\hbar^4$ term. The $t_k$ for higher $k$ begin to appear at successively higher orders starting at $\hbar^6$ with $t_2$~\cite{Johnson:2020heh}. The other simplification is that the expansion is simply in inverse powers of $x$, and because $u_0(x)$ vanishes, the resulting integrands are of this form also. Therefore, results can be obtained rather swiftly.

The  spectral density is $\rho(E){=}(\pi\hbar)^{-1}{\rm Im}\int^\mu {\widehat R}(x,E)dx$, and so it can be seen that the first term in Eq.~(\ref{eq:gelfand-dikii-A}) gives  the crucial hard wall contribution discussed in the main text. It is the piece of the leading spectral density~(\ref{eq:spectral-density-leading}) that comes from integrating over the $x>0$ region.  Putting $u_0(x)$ and its derivatives to zero, and inserting the results for $u_2(x)$ and  $u_4(x)$ gives:
\bea
\label{eq:resolvent-results}
\int^\mu {\widehat R}_1(x,E)dx &=& -\frac{1}{4}\left(\Gamma^2-\frac14\right)\frac{1}{(-E)^{3/2}}\frac{1}{\mu}\ , \nonumber\\
\int^\mu {\widehat R}_2(x, E)dx &=& \frac{1}{16}\left(\Gamma^2-\frac14\right)\left(\Gamma^2-\frac94\right)\left(\frac{2t_1}{(-E)^{3/2}\mu^4}+\frac{1}{(-E)^{5/2}\mu^3}\right)\ .
\eea
As a test against known results,  the (time non-reversible) case of ${\cal N}{=}1$ JT supergravity treated in  Section 5.2.2 of ref.~\cite{Stanford:2019vob} is in fact $\Gamma{=}0$, and the $t_k$ decomposition was worked out in ref.~\cite{Johnson:2020heh} as $t_k {=} \pi^{2k}/(k!)^2$, so  $t_1{=}\pi^2$. Also, $\mu{=}1$ for this case. Putting those into Eqs.~(\ref{eq:resolvent-results}) and writing $z_1^2{=}{-}E$ yields:
\be
\int^\mu {\widehat R}_1(x,E)dx \quad \longrightarrow \quad \frac{1}{2^4}\frac{1}{z_1^3}\ ,\quad 
\int^\mu {\widehat R}_2(x, E)dx \quad \longrightarrow \quad\frac{1}{2^8}\left(\frac{18\pi^2}{z_1^3}+\frac{9}{z_1^5}\right)\ ,
\ee
which are (up to a $\sqrt{2}$ factor)  ref.~\cite{Stanford:2019vob}'s $R_g{\rm SJT}(-z_1^2)$,   obtained from the loop equations/recurrence, so this approach nicely matches on to supersymmetric JT gravity results.

Denoting $\omega_{g,1}^{(b)}(z_1){=}
2z_1\int^\mu {\widehat R}(x,E)$,  
 the following convention  will be used for the relation between volumes and resolvent-type quantities (which differs slightly from those used in refs.~\cite{Stanford:2019vob,Collier:2023cyw}):
\be
\omega_{g,1}^{(b)}(z_1) = \int_0^\infty {\widehat V}_{g,1}^{(b)}(P_1)  {\rm e}^{-z_1P_1} 2P_1 dP_1\ ,
\ee
(where $P_1$ would be replaced by the geodesic length $b_1$ in the JT supergravity case).
Using this, the   volumes computed here to this order are, generally:
\be
{\widehat V}_{1,1}^{(b)}(P_1) = -\left(\Gamma^2-\frac14\right)\left(\frac{1}{2\mu}\right)\ , \quad 
{\widehat V}_{2,1}^{(b)}(P_1) = \frac{1}{96\mu^3}\left(\Gamma^2 -\frac14\right)\left(\Gamma^2 -\frac94\right)\left(P_1^2+\frac{12 t_1}{\mu}\right)\ .
\ee
A check shows that $\Gamma{=}0$, $t_1{=}\pi^2$ and $\mu{=}1$ (and sending $P_1{\to} b_1$) gives $V_{1}{=}1/8$ and $V_{2}{=}3{\times} 2^{-9}(b_1^2+12\pi^2)$ of ref.~\cite{Stanford:2019vob}.

Finally, the new quantum volumes for the supersymmetric Virasoro string models of this paper can be computed by using (from Eq.~(\ref{eq:new-teekay})) 
\be
t_1=2\sqrt{2}\pi^3(Q^2+{\widehat Q}^2)=4\sqrt{2}\pi^3\left(\frac{c-\frac{15}{2}}{3}\right)\ ,\quad {\rm and }\quad \mu = 4\sqrt{2}\pi\ ,
\ee
 which in the case  $\Gamma{=}0$ gives:
\be
\omega^{(b)}_{1,1}(z_1) = \frac{\sqrt{2}}{2^6\pi}\frac{1}{z_1^2}\ ,\quad \omega^{(b)}_{2,1}(z_1) = \frac{\sqrt{2}}{2^{15}\pi^3}\left(\frac{6\pi^2\left(c-\frac{15}{2}\right)}{z_1^2}+\frac{9}{z_1^4}\right)\ ,
\ee
and hence
\be
{\widehat V}^{(b)}_{1,1}(z_1) = \frac{\sqrt{2}}{2^6\pi}\ ,\quad 
{\widehat V}^{(b)}_{2,1}(z_1) = \frac{3\sqrt{2}}{2^{16}\pi^3}\left(P_1^2 + 4\pi^2\left(c-\frac{15}{2}\right)\right)\ .
\ee
 Interestingly, the central charge appears naturally here through~$t_1$ (analogous to what happens in the regular VMS case~\cite{Johnson:2024bue}, but through a very different perturbative expansion) and appears at $g{=}2$ instead of $g{=}1$.

Further recursive solving of the two ODEs~(\ref{eq:big-string-equation}) and~(\ref{eq:gelfand-dikii-A}) yields the higher ${\widehat V}^{(b)}_{g,1}$, a process deftly playing the role of topological recursion~\cite{Eynard:2014zxa} in this formalism. As a last example, the next order is, in general form:
\be
{\widehat V}^{(b)}_{3,1}(z_1) =-\frac{\left(\Gamma^2-\frac14\right)\left(\Gamma^2-\frac94\right)}{1920\mu^5}
\left\{   
\left(\Gamma^2-\frac{25}{4}\right)P_1^4 
+\left(\Gamma^2-\frac{21}{4}\right)\frac{80t_1}{\mu}P_1^2
+\left(\Gamma^2-\frac{21}{4}\right)\frac{960t_1^2}{\mu^2}-\left(\Gamma^2-\frac{25}{4}\right)\frac{320t_2}{\mu}
\right\}\ ,
\ee
following from the fact that the next order in the expansion for $u(x)$ is~\cite{Johnson:2020heh}:
\be
u_6(x)=\hbar^6\left(\Gamma^2-\frac14\right)\left(\Gamma^2-\frac94\right)\left[
\frac{7t_1^2}{x^8}\left(\Gamma^2-\frac{21}{4}\right)
-\frac{2t_2}{x^7}\left(\Gamma^2-\frac{25}{4}\right)
\right]\ ,
\ee
and for the super volumes, there is the additional relation from Eq.~(\ref{eq:new-teekay}) that $t_2{=}2\sqrt{2}\frac{\pi^5}{4}\left(Q^4+{\widehat Q}^4\right)$, which can be written in terms of the central charge $c$. Note that in ${\cal N}{=}1$ JT supergravity this  (upon using $\Gamma{=}0$, $t_2{=}\pi^4/4$, $\mu{=}1$, and setting $P_1{\to} b_1$) becomes $V_{3,1}(b_1){=} 3{\times}2^{-13}[5b_1^4+336\pi^2 b_1^2+3632\pi^4]$ (matching the corresponding entry in {\it e.g.,}  the Appendices of ref.~\cite{Fuji:2023wcx}).

Note that the formalism and matrix model definition of this paper naturally yields  more general behaviour incorporating other $\Gamma$. As discussed in the main text, the cases $\Gamma{=}{\pm}\frac12$ are special, in that the volumes vanish, as do various multi-point correlators at leading order and higher genus. (The other evident special points for half-integer $\Gamma$ also deserve investigation.)


\bibliographystyle{apsrev4-1}
\bibliography{Fredholm_super_JT_gravity1,Fredholm_super_JT_gravity2}

\end{document}